\documentclass[11pt]{article}
\usepackage{amssymb}
\usepackage{amsfonts}
\usepackage{amsthm}
\newtheorem{prop}{Proposition}
\newtheorem*{lemma}{Lemma}
\newcommand{\abs}[1]{\mbox{$|#1|$}}
\newcommand{\norm}[1]{\mbox{$\| #1\|$}}
\newcommand{\alg}[1]{\mbox{$\mathfrak{#1}$}}
\newcommand{\hil}[1]{\mbox{$\mathcal{#1}$}}

\newcommand{\twoqubit}{\mbox{$\mathbf{B}(\mathbb{C}^{n})\otimes{\mathbf{B}(\mathbb{C}}^{n})$}}

\title{Generic Bell correlation between arbitrary local algebras in quantum field theory}
\author{Hans Halvorson\thanks{\tt{hphst1+@pitt.edu}} \\ {\normalsize
    Departments of Mathematics and Philosophy } \\ {\normalsize University of Pittsburgh}  
\and Rob Clifton\thanks{\tt{rclifton+@pitt.edu}} \\ {\normalsize Departments of
Philosophy and History \& Philosophy of Science}  \\ {\normalsize
University of Pittsburgh} }
\date{November 4, 1999}
\begin{document}
\maketitle 
\begin{abstract}  We prove that for any two commuting von Neumann
  algebras of infinite type, the open set of Bell correlated states
  for the two algebras is norm dense.  We then apply this result to
  algebraic quantum field theory --- where all local algebras are of
  infinite type --- in order to show that for any two spacelike
  separated regions, there is an open dense set of field states that
  dictate Bell correlations between the regions.  We also show that
  any vector state cyclic for one of a pair of commuting nonabelian
  von Neumann algebras is entangled (i.e., nonseparable) across the
  algebras---from which it follows that every field state with bounded
  energy is entangled across any two spacelike separated regions.
\end{abstract}  
%\newpage
\section{Introduction}
There are many senses in which the phenomenon of Bell correlation, originally discovered and
investigated in the context of elementary nonrelativistic quantum mechanics~\cite{bell,chsh}, is
``generic'' in quantum field theory models.  For example, it has been shown that every pair of
commuting nonabelian von Neumann algebras possesses \emph{some} normal state with maximal
Bell correlation~\cite{review} (see also~\cite{violator}).  Moreover, in most standard quantum
field models, \emph{all} normal states are maximally Bell correlated across spacelike separated
tangent wedges or double cones~[3,5--8].  Finally, every bounded energy state in quantum field
theory sustains maximal Einstein-Podolsky-Rosen correlations across arbitrary spacelike
separated regions~\cite{redhead}, and has a form of nonlocality that may be evinced by means of
the state's violation of a conditional Bell inequality~\cite{conditional}.  (We also note that the
study of Bell correlation in quantum field theory has recently borne fruit in the introduction of a
new algebraic invariant for an inclusion of von Neumann algebras~\cite{invariants,structure}.)

Despite these numerous results, it remains an open question whether ``most'' states will have some
or other Bell correlation relative to \emph{arbitrary} spacelike separated regions.  Our main
purpose in this note is to verify that this is so: for any two spacelike separated regions, there is an
open dense set of states which have Bell correlations across those two regions.  

In section~II we prove the general result that for any pair of mutually commuting von Neumann
algebras of infinite type, a dense set of vectors will induce states which are Bell correlated across
these two algebras.  In section~III we introduce, following~\cite{werner}, a notion of
``nonseparability'' of states that generalizes, to mixed states, the idea of an entangled pure state
vector.  We then show that for a pair of nonabelian von Neumann algebras, a vector cyclic for
either algebra induces a nonseparable state.  Finally, in section~IV we apply these results to
algebraic quantum field theory.

\section{Bell correlation between infinite von Neumann algebras}
Let $\hil{H}$ be a Hilbert space, let $\hil{S}$ denote the set of unit
vectors in $\hil{H}$, and let $\mathbf{B}(\hil{H})$ denote the set of
bounded linear operators on $\hil{H}$.  We will use the same notation
for a projection in $\mathbf{B}(\hil{H})$ and for the subspace in
$\hil{H}$ onto which it projects.  If $x\in \hil{S}$, we let $\omega
_{x}$ denote the state of $\mathbf{B}(\hil{H})$ induced by $x$.  Let
$\alg{R}_{1},\alg{R}_{2}$ be von Neumann algebras acting on $\hil{H}$
such that $\alg{R}_{1}\subseteq \alg{R}'_{2}$, and let $\alg{R}_{12}$
denote the von Neumann algebra $\{ \alg{R}_{1}\cup \alg{R}_{2} \} ''$
generated by $\alg{R}_{1}$ and $\alg{R}_{2}$.
Following~\cite{invariants}, we set
\begin{eqnarray}
\hil{T}_{12} &\equiv & \Bigl\{ (1/2)
[A_{1}(B_{1}+B_{2})+A_{2}(B_{1}-B_{2})] : \nonumber \\
& & \quad A_{i}=A_{i}^{*}\in \alg{R}_{1}, B_{i}=B_{i}^{*}\in \alg{R}_{2}, -I\leq
A_{i},B_{i} \leq I \Bigr\} .\end{eqnarray}
Elements of $\hil{T}_{12}$ are called \emph{Bell operators} for
$\alg{R}_{12}$.  For a given state $\omega$ of $\alg{R}_{12}$, let
\begin{equation}
\beta (\omega )\equiv \sup \{ \abs{\omega (R) }:R\in \hil{T}_{12} \}
\label{beta}
.\end{equation}
If $\omega =\omega _{x}|_{\alg{R}_{12}}$ for some $x\in \hil{S}$, we write
$\beta (x)$ to
abbreviate $\beta (\omega _{x}|_{\alg{R}_{12}})$.  From~(\ref{beta}), it
follows that the map
$\omega \rightarrow \beta (\omega )$ is norm continuous from the state
space of $\alg{R}_{12}$
into $[1,\sqrt{2}]$~\cite[Lemma~2.1]{invariants}.  Since the map
$x\rightarrow \omega
_{x}|_{\alg{R}_{12}}$ is continuous from $\hil{S}$, in the vector norm
topology, into the
(normal) state space of $\alg{R}_{12}$, in the norm topology, it also
follows that $x\rightarrow
\beta (x)$ is continuous from $\hil{S}$ into $[1,\sqrt{2}]$.  If $\beta
(\omega )>1$, we say that
$\omega$ violates a Bell inequality, or is \emph{Bell correlated}.  In this
context, Bell's
theorem~\cite{bell} is the statement that a local hidden variable model of
the correlations that
$\omega$ dictates between $\alg{R}_{1}$ and $\alg{R}_{2}$ is only possible
if $\beta (\omega
)=1$.  Note that the set of states $\omega$ on $\alg{R}_{12}$ that violate
a Bell inequality is open (in the
norm topology) and, similarly, the set of vectors $x\in \hil{S}$ that induce Bell
correlated states on
$\alg{R}_{12}$ is open (in the vector norm topology).

We assume now that the pair $\alg{R}_{1},\alg{R}_{2}$ satisfies the
\emph{Schlieder property}.  That
is, if $A\in \alg{R}_{1}$ and $B\in \alg{R}_{2}$ such that $AB = 0$, then
either $A=0$ or
$B=0$.  Let $V\in \alg{R}_{1}$ and $W\in \alg{R}_{2}$ be nonzero partial
isometries.  Suppose
that the
initial space $V^{*}V$ of $V$ is orthogonal to the final space $VV^{*}$
of $V$;
or, equivalently,
that $V^{2}=0$.  Similarly, suppose $W^{2}=0$.
Consider the projections
\begin{equation}
 E=V^{*}V+VV^{*},\ \ F=W^{*}W+WW^{*}.
 \end{equation}
   We show that there is a Bell operator $\widetilde{R}$ for
$\alg{R}_{12}$ such that $\widetilde{R}y=\sqrt{2}y$ for some unit vector
$y\in EF$, and
$\widetilde{R}(I-E)(I-F)=(I-E)(I-F)$.

Let \begin{equation} \begin{array}{ll}
A_{1}=V+V^{*}  &\qquad B_{1}=W+W^{*} \\
A_{2}=i(V^{*}-V) &\qquad B_{2}=i(W^{*}-W) \\
A_{3}=[V,V^{*}]  &\qquad B_{3}=[W,W^{*}] .\end{array} \end{equation}
Note that $A_{i}^{2}=E$, the $A_{i}$ are self-adjoint contractions in
$\alg{R}_{1}$,
$A_{i}E=EA_{i}=A_{i}$, and $[A_{1},A_{2}]=2iA_{3}$.  Similarly,
$B_{i}^{2}=F$, the
$B_{i}$ are self-adjoint contractions in $\alg{R}_{2}$,
$B_{i}F=FB_{i}=B_{i}$, and
$[B_{1},B_{2}]=2iB_{3}$.  If we let $R$ denote the Bell operator
constructed from
$A_{i},B_{i}$, a straightforward calculation shows that~(cf.~\cite{violator})
\begin{equation}
R^{2}=EF-\frac{1}{4}[A_{1},A_{2}][B_{1},B_{2}] =EF+A_{3}B_{3} . \label{square}
\end{equation}
Note that $P\equiv VV^{*}\not=0$ is the spectral projection for $A_{3}$ corresponding to
eigenvalue~$1$, and $Q\equiv WW^{*}\not=0$ is the spectral projection for $B_{3}$
corresponding to eigenvalue~$1$.  Since $\alg{R}_{1},\alg{R}_{2}$ satisfy the Schlieder
property, there is a unit vector $y\in PQ$, and thus $A_{3}B_{3}y=y$.  Since $PQ<EF$, it
follows from~(\ref{square}) that $R^{2}y=2y$.  Thus, we may assume without loss of generality
that $Ry=\sqrt{2}y$.  (If $Ry\neq \sqrt{2}y$, then interchange $B_{1},B_{2}$ and replace
$A_{1}$ with $-A_{1}$.  Note that the resulting Bell operator $R'=-R$ and
$R'y_{0}=\sqrt{2}y_{0}$, where $y_{0}\equiv (\sqrt{2}y-Ry)/\norm{\sqrt{2}y-Ry}\in EF$.)

Now for $i=1,2$, let $\widetilde{A}_{i}=(I-E)+A_{i}$ and
$\widetilde{B}_{i}=(I-F)+B_{i}$.  It is easy to see that
$\widetilde{A}_{i}^{2}=I$ and $\widetilde{B}_{i}^{2}=I$, so the
$\widetilde{A}_{i}$ and $\widetilde{B}_{i}$ are again self-adjoint
contractions in $\alg{R}_{1}$ and $\alg{R}_{2}$ respectively.  If we
let $\widetilde{R}$ denote the corresponding Bell operator, a
straightforward calculation shows that
\begin{equation}
\widetilde{R}=(I-E)(I-F)+(I-E)B_{1}+A_{1}(I-F)+R .\end{equation}
Since the $\sqrt{2}$ eigenvector $y$ for $R$ lies in $EF$, we have
$\widetilde{R}y=Ry=\sqrt{2}y$.
Furthermore, since $A_{i}(I-E)=0$ and $B_{i}(I-F)=0$, we have
$\widetilde{R}(I-E)(I-F)=(I-E)(I-F)$ as required.

A special case of the following result, where $\alg{R}_{1}$ and $\alg{R}_{2}$ are type
I$_{\infty}$ factors, was proved as \cite[Prop.~1]{genbell}.  Recall that $\alg{R}$ is said to be
\emph{of infinite type} just in case the identity $I$ is equivalent, in $\alg{R}$, to one of its
proper subprojections.

\begin{prop} Let $\alg{R}_{1},\alg{R}_{2}$ be von Neumann
  algebras acting on $\hil{H}$ such that $\alg{R}_{1}\subseteq
  \alg{R}'_{2}$, and $\alg{R}_{1},\alg{R}_{2}$ satisfy the Schlieder
  property.  If $\alg{R}_{1},\alg{R}_{2}$ are of infinite type, then
  there is an open dense subset of vectors in $\hil{S}$ which induce Bell correlated states for
$\alg{R}_{12}$. \label{inf} \end{prop}

Note that the hypotheses of this proposition are invariant under isomorphisms of $\alg{R}_{12}$.
Thus, by making use of the universal normal representation of
$\alg{R}_{12}$~\cite[p.~458]{kad}, in which all normal states are vector states, it follows that
the set of states Bell correlated for $\alg{R}_{1},\alg{R}_{2}$ is norm dense in the normal state
space of $\alg{R}_{12}$.  

\begin{proof}[Proof of the proposition:]  Since $\alg{R}_{1}$ is infinite,
  there is a properly infinite projection $P\in
  \alg{R}_{1}$~\cite[Prop.~6.3.7]{kad}.  Since $P$ is properly
  infinite, we may apply the halving lemma~\cite[Lemma~6.3.3]{kad}
  repeatedly to obtain a countably infinite family $\{ P_{n} \}$ of
  mutually orthogonal projections such that $P_{n}\sim P_{n+1}$ for
  all $n$ and $\sum _{n=1}^{\infty} P_{n}=P$.  (Halve $P$ as
  $P_{1}+F_{1}$; then halve $F_{1}$ as $P_{2}+F_{2}$, and so on.  Now
  replace $P_{1}$ by $P-\sum _{n=2}^{\infty}P_{n}$; cf.
  \cite[Lemma~6.3.4]{kad}.)  Let $P_{0}\equiv I-P$.  For each $n\in
  \mathbb{N}$, let $V_{n}$ denote the partial isometry with initial
  space $V_{n}^{*}V_{n}=P_{n}$ and final space
  $V_{n}V_{n}^{*}=P_{n+1}$.  By the same reasoning, there is a
  countable family $\{ Q_{n} \}$ of mutually orthogonal projections in
  $\alg{R}_{2}$ and partial isometries $W_{n}$ with
  $W^{*}_{n}W_{n}=Q_{n}$ and $W_{n}W^{*}_{n}=Q_{n+1}$.  For each $n
  \in \mathbb{N}$, let \begin{equation}
\begin{array}{ll}
A_{1,n}=V_{n+1}+V_{n+1}^{*}    &\qquad B_{1,n}=W_{n+1}+W_{n+1}^{*}, \\
A_{2,n}=i(V_{n+1}^{*}-V_{n+1}) &\qquad B_{2,n}=i(W_{n+1}^{*}-W_{n+1}),
\end{array} \end{equation}
and let \begin{equation}
\begin{array}{l}
E_{n} = V_{n+1}^{*}V_{n+1}+V_{n+1}V_{n+1}^{*}  =  P_{n+1}+P_{n+2} , \\
F_{n} =W^{*}_{n+1}W_{n+1}+W_{n+1}W_{n+1}^{*}=Q_{n+1}+Q_{n+2}.
\end{array}  \end{equation}
Define $\widetilde{A}_{i,n}$ and $\widetilde{B}_{i,n}$ as in the discussion
preceding this
proposition, let $\widetilde{R}_{n}$ be the corresponding Bell operator,
and let the unit vector
$y_{n}\in E_{n}F_{n}$ be the $\sqrt{2}$ eigenvector for $\widetilde{R}_{n}$.

Now, let $x$ be any unit vector in $\hil{H}$.  Since $\sum
_{i=0}^{n}P_{i}\leq I-E_{n}$, we
have $(I-E_{n})\rightarrow I$ in the strong-operator topology.  Similarly,
$(I-F_{n})\rightarrow
I$ in the strong-operator topology.  Therefore if we let \begin{equation}
x_{n}\equiv \frac{(I-E_{n})(I-F_{n})x}{\norm{(I-E_{n})(I-F_{n})x}} , \end{equation}
we have \begin{equation}
x =\lim _{n}(I-E_{n})(I-F_{n})x=\lim _{n}x_{n} .\end{equation}
Note that the inner product $\langle x_{n},y_{n} \rangle =0$, and thus
\begin{equation}
z_{n}\equiv (1-n^{-1})^{1/2}x_{n}+n^{-1/2}y_{n} \end{equation}
is a unit vector for all $n$.  Since $\lim _{n}z_{n}=x$, it suffices
to observe that each $z_{n}$ is
Bell correlated for $\alg{R}_{12}$.  Recall that
$\widetilde{R}_{n}(I-E_{n})(I-F_{n})=(I-E_{n})(I-F_{n})$, and thus
$\widetilde{R}_{n}x_{n}=x_{n}$.  A simple calculation then reveals
that \begin{equation}
\beta (z_{n})\:\geq \:\langle \widetilde{R}_{n}z_{n},z_{n} \rangle
\:=\:(1-n^{-1})+n^{-1}\sqrt{2}\:>\:1 .\end{equation} 
\end{proof}

\section{Cyclic vectors and entangled states}
Proposition~\ref{inf} establishes that Bell correlation is generic for
commuting pairs of infinite von Neumann algebras.  However, we are
given no information about the character of the correlations of
particular states.  We provide a partial remedy for this in the next
proposition, where we show that any vector cyclic for $\alg{R}_{1}$
(or for $\alg{R}_{2}$) induces a state that is not classically
correlated; i.e., it is ``nonseparable.''

Again, let $\alg{R}_{1},\alg{R}_{2}$ be von Neumann algebras on
$\hil{H}$ such that $\alg{R}_{1}\subseteq \alg{R}'_{2}$.  Recall that
a state $\omega$ of $\alg{R}_{12}$ is called a \emph{normal product
  state} just in case $\omega$ is normal, and there are states $\omega _{1}$ of
$\alg{R}_{1}$ and $\omega _{2}$ of $\alg{R}_{2}$ such that
\begin{equation}
\omega (AB)=\omega _{1}(A)\omega _{2}(B) ,\end{equation}
for all $A\in \alg{R}_{1},B\in \alg{R}_{2}$.  Werner~\cite{werner}, in
dealing with the case of $\twoqubit$, defined a
density operator $D$ to be \emph{classically correlated} --- the term \emph{separable} is
now more commonly used --- just in case
$D$ can be approximated in trace norm by convex combinations of density
operators of form $D_{1}\otimes D_{2}$.  Although Werner's definition
of nonseparable states directly generalizes
the traditional notion of pure entangled states, he showed that a
nonseparable mixed state need
not violate a Bell inequality; thus, Bell correlation is in general a
sufficient, though not necessary
condition for a state's being nonseparable.  On the other hand, it has
since been shown that
nonseparable states often possess more subtle forms of nonlocality, which may be
indicated by measurements more general than the single ideal measurements
which can indicate Bell correlation~\cite{popescu}.  (See \cite{genbell,gensep} for
further discussion.)

In terms of the linear functional representation of states, Werner's
separable states are those in the norm closed convex hull of the
product states of $\twoqubit$.  However, in case of the more general
setup --- i.e., $\alg{R}_{1}\subseteq \alg{R}'_{2}$, where
$\alg{R}_{1},\alg{R}_{2}$ are arbitrary von Neumann algebras on
$\hil{H}$ --- the choice of topology on the normal state space of
$\alg{R}_{12}$ will yield in general different definitions of
separability.  Moreover, it has been argued that norm convergence of a
sequence of states can never be verified in the laboratory, and as a
result, the appropriate notion of physical approximation is given by
the (weaker) weak-$*$ topology~\cite{emc:alg,haag}.  And the weak-$*$
and norm topologies do not generally coincide \emph{even} on the
normal state space~\cite{del:ont}.

For the next proposition, then, we will suppose that the separable
states of $\alg{R}_{12}$ are those normal states in the weak-$*$
closed convex hull of the normal product states.  Note that $\beta
(\omega )=1$ if $\omega$ is a product state, and since $\beta $ is a
convex function on the state space, $\beta (\omega )=1$ if $\omega$ is
a convex combination of product states~\cite[Lemma~2.1]{invariants}.
Furthermore, since $\beta $ is lower semicontinuous in the weak-$*$
topology~\cite[Lemma~2.1]{invariants}, $\beta (\omega )=1$ for any
separable state.  Conversely, any Bell correlated state must be
nonseparable.

We now introduce some notation that will aid us in the proof of our result.
For a state $\omega $ of the von Neumann algebra $\alg{R}$ and an operator
$A\in \alg{R}$, define the state $\omega ^{A}$ on $\alg{R}$ by \begin{equation}
\omega ^{A}(X)\equiv \frac{\omega (A^{*}XA)}{\omega (A^{*}A)} ,\end{equation}
if $\omega (A^{*}A) \neq 0$, and let $\omega ^{A}=\omega$ otherwise.
Suppose now that
$\omega (A^{*}A)\neq 0$ and $\omega $ is a convex combination of states:
\begin{equation}
\omega =\sum _{i=1}^{n}\lambda _{i}\omega _{i} .\end{equation}  Then, letting
$\lambda ^{A}_{i}\equiv \omega (A^{*}A)^{-1}\omega _{i}(A^{*}A)\lambda _{i}$,
$\omega ^{A}$ is again a convex combination 
\begin{equation}
\omega ^{A}=\sum _{i=1}^{n}\lambda ^{A}_{i}\omega _{i}^{A} . \end{equation}
Moreover, it is not difficult to see that the map $\omega \rightarrow \omega
^{A}$ is weak-$*$ continuous at any point $\rho$ such that $\rho
(A^{*}A)\neq 0$.  Indeed, let
$\hil{O}_{1}=N(\rho ^{A}:X_{1},\dots ,X_{n},\epsilon )$ be a weak-$*$
neighborhood
of $\rho ^{A}$.  Then, taking $\hil{O}_{2}=N(\rho :A^{*}A,A^{*}X_{1}A,\dots
,A^{*}X_{n}A ,\delta )$ and $\omega \in \hil{O}_{2}$, we have
\begin{equation}
\abs{ \rho (A^{*}X_{i}A) -\omega (A^{*}X_{i}A) } <\delta   , \end{equation}
for $i=1,\dots
,n$, and \begin{equation}
\abs{ \rho (A^{*}A) -\omega (A^{*}A) } <\delta   . \end{equation}  By
choosing $\delta < \rho
(A^{*}A)\neq 0$, we also have $\omega (A^{*}A)\neq 0$, and
thus \begin{equation} \abs{ \rho ^{A}(X_{i})-\omega ^{A}(X_{i}) }< O(\delta
)\leq \epsilon
,\end{equation}
for an appropriate choice of $\delta$.  That is,  $\omega ^{A}\in
\hil{O}_{1}$ for all $\omega \in
\hil{O}_{2}$ and $\omega \rightarrow \omega ^{A}$ is weak-$*$ continuous at
$\rho$.

Specializing to the case where $\alg{R}_{1}\subseteq \alg{R}'_{2}$, and
$\alg{R}_{12} =\{
\alg{R}_{1} \cup \alg{R}_{2} \} ''$, it is clear from the above that for
any normal product state
$\omega$ of $\alg{R}_{12}$ and for $A\in \alg{R}_{1}$, $\omega ^{A}$ is
again a normal
product state.  The same is true if $\omega$ is a convex combination of
normal product states, or
the weak-$*$ limit of such combinations.  We summarize the results of
this discussion in
the following lemma:

\begin{lemma} For any separable state $\omega$ of $\alg{R}_{12}$ and any $A\in
\alg{R}_{1}$, $\omega ^{A}$ is again separable.  \end{lemma}

\begin{prop}  Let $\alg{R}_{1},\alg{R}_{2}$ be nonabelian von Neumann
  algebras such that $\alg{R}_{1}\subseteq \alg{R}'_{2}$.  If $x$ is
  cyclic for $\alg{R}_{1}$, then $\omega _{x}$ is nonseparable across
  $\alg{R}_{12}$.  \label{cyclic} \end{prop}

\begin{proof} From~\cite[Lemma~2.1]{invariants}, there is a normal state $\rho$ of
  $\alg{R}_{12}$ such that $\beta (\rho )=\sqrt{2}$.  But since all
  normal states are in the (norm) closed convex hull of vector
  states~\cite[Thm~7.1.12]{kad}, and since $\beta$ is norm continuous
  and convex, there is a vector $v\in \hil{S}$ such that $\beta
  (v)>1$.  By the continuity of $\beta$ (on $\hil{S}$), there is an
  open neighborhood $\hil{O}$ of $v$ in $\hil{S}$ such that $\beta (y)
  >1$ for all $y\in \hil{O}$.  Since $x$ is cyclic for $\alg{R}_{1}$,
  there is an $A\in \alg{R}_{1}$ such that $Ax \in \hil{O}$.  Thus,
  $\beta (Ax)>1$ which entails that $\omega _{Ax}=(\omega _{x})^{A}$
  is a nonseparable state for $\alg{R}_{12}$.  This, by the preceding
  lemma, entails that $\omega _{x}$ is nonseparable.
\end{proof}

Note that if $\alg{R}_{1}$ has at least one cyclic vector $x\in \hil{S}$, then $\alg{R}_{1}$ has
a dense set of cyclic vectors in $\hil{S}$~\cite{dix:vec}.  Since each of the corresponding vector
states is nonseparable across $\alg{R}_{12}$, Proposition~\ref{cyclic} shows that if
$\alg{R}_{1}$ has a cyclic vector, then the (open) set of vectors inducing nonseparable states
across $\alg{R}_{12}$ is dense in $\hil{S}$.  On the other hand, since the existence of a cyclic
vector for $\alg{R}_{1}$ is not invariant under isomorphisms of $\alg{R}_{12}$,
Proposition~\ref{cyclic} does not entail that if $\alg{R}_{1}$ has a cyclic vector, then there is a
norm dense set of nonseparable states in the entire normal state space of $\alg{R}_{12}$.  (Cf.
the analogous discussion preceding the proof of Proposition~\ref{inf}.)  Indeed, if we let
$\alg{R}_{1}=\mathbf{B}(\mathbb{C}^{2})\otimes I$, $\alg{R}_{2}=I\otimes
\mathbf{B}(\mathbb{C}^{2})$, then any entangled state vector is cyclic for $\alg{R}_{1}$; but,
the set of nonseparable states of $\mathbf{B}(\mathbb{C}^{2})\otimes
\mathbf{B}(\mathbb{C}^{2})$ is \emph{not} norm dense~\cite{gensep,z-guy}.  However, if in
addition to $\alg{R}_{1}$ or $\alg{R}_{2}$ having a cyclic vector, $\alg{R}_{12}$ has a
separating vector (as is often the case in quantum field theory), then all normal states of
$\alg{R}_{12}$ are vector states~\cite[Thm.~7.2.3]{kad}, and it follows that the nonseparable
states \emph{will} be norm dense in the entire normal state space of $\alg{R}_{12}$.

\section{Applications to algebraic quantum field theory}
Let $(M,g)$ be a relativistic spacetime and let $\alg{A}$ be a unital $C^{*}$-algebra.
The basic mathematical object of algebraic quantum field theory
(see~\cite{haag,borchers,dimock}) is an association between precompact open subsets $O$ of
$M$ and $C^{*}$-subalgebras $\alg{A}(O)$ of $\alg{A}$.  (We assume that each
$\alg{A}(O)$ contains the identity $I$ of $\alg{A}$.)  The motivation for this association is the
idea that $\alg{A}(O)$ represents observables that can be measured in the region $O$.  With
this in mind, one assumes \begin{enumerate}  
\item \emph{Isotony:}  If $O_{1}\subseteq O_{2}$, then $\alg{A}(O_{1})\subseteq
\alg{A}(O_{2})$. 
\item \emph{Microcausality:}  $\alg{A}(O')\subseteq \alg{A}(O)'$.   
\end{enumerate}
Here $O'$ denotes the interior of the set of all points of $M$ that are spacelike to every point in
$O$.  

In the case where $(M,g)$ is Minkowski spacetime, it is assumed in addition that there is a
faithful representation $\mathbf{x}\rightarrow \alpha _{\mathbf{x}}$ of the translation group of
$M$ in the group of automorphisms of $\alg{A}$ such that \begin{enumerate} 
\item[3.] \emph{Translation Covariance:}  $\alpha
_{\mathbf{x}}(\alg{A}(O))=\alg{A}(O+\mathbf{x})$.  
\item[4.] \emph{Weak Additivity:} For any $O\subseteq M$, $\alg{A}$ is the smallest
$C^{*}$-algebra containing $\cup _{\mathbf{x}\in M}\alg{A}(O+\mathbf{x})$.
\end{enumerate}

The class of physically relevant representations of $\alg{A}$ is decided by further desiderata such
as --- in the case of Minkowski spacetime --- a unitary representation of the group of translation
automorphisms which satisfies the spectrum condition.  Relative to a fixed representation $\pi$,
we let $\alg{R}_{\pi}(O)$ denote the von Neumann algebra $\pi (\alg{A}(O))''$ on the
representation space $\hil{H}_{\pi}$.  In what follows, we consider only nontrivial
representations (i.e., $\dim \hil{H}_{\pi}>1$), and we let $\hil{S}_{\pi}$ denote the set of unit
vectors in $\hil{H}_{\pi}$.       

\begin{prop}  Let $\{ \alg{A}(O)\}$ be a net of local algebras over Minkowski spacetime.  Let
$\pi$ be any representation in the local quasiequivalence class of some irreducible vacuum
representation (e.g. superselection sectors in the sense of Doplicher-Haag-Roberts~\cite{dhr} or
Buchholz-Fredenhagen~\cite{bf}).  If $O_{1},O_{2}$ are any two open subsets of $M$ such
that $O_{1}\subseteq O_{2}'$, then the set of vectors inducing Bell correlated states for
$\alg{R}_{\pi}(O_{1}),\alg{R}_{\pi}(O_{2})$ is open and dense~in~$\hil{S}_{\pi}$. 
\label{irred}  \end{prop} 

\begin{proof} Let $O_{3},O_{4}$ be precompact open subsets of $M$ such that
$O_{3}\subseteq O_{1}, O_{4}\subseteq O_{2}$, and such that $O_{3}+N\subseteq O_{4}'$
for some neighborhood $N$ of the origin.  In an irreducible vacuum representation $\phi$, local
algebras are of infinite type~\cite[Prop.~1.3.9]{hor:int}, and since $O_{3}+N\subseteq O_{4}'$,
the Schlieder property holds for
$\alg{R}_{\phi}(O_{3}),\alg{R}_{\phi}(O_{4})$~\cite{sch:ein}.  If $\pi$ is any representation
in the local quasiequivalence class of $\phi$, these properties hold for
$\alg{R}_{\pi}(O_{3}),\alg{R}_{\pi}(O_{4})$ as well.  Thus, we may apply Proposition~1 to
conclude that the set of vectors inducing Bell correlated states for
$\alg{R}_{\pi}(O_{3}),\alg{R}_{\pi}(O_{4})$ is dense in $\hil{S}_{\pi}$.  Finally, note that
any state Bell correlated for $\alg{R}_{\pi}(O_{3}),\alg{R}_{\pi}(O_{4})$ is Bell correlated for
$\alg{R}_{\pi}(O_{1}),\alg{R}_{\pi}(O_{2})$.  \end{proof}

\begin{prop} Let $(M,g)$ be a globally hyperbolic spacetime, let $\{ \alg{A}(O)\}$ be the net of
local observable algebras associated with the free Klein-Gordon field~\cite{dimock}, and let $\pi$
be the GNS representation of some quasifree Hadamard state~\cite{wald}.  If $O_{1},O_{2}$
are any two open subsets of $M$ such that $O_{1}\subseteq O_{2}'$, then the set of vectors
inducing Bell correlated states for $\alg{R}_{\pi}(O_{1}),\alg{R}_{\pi}(O_{2})$ is open and
dense in $\hil{S}_{\pi}$.  \end{prop}     

\begin{proof} The regular diamonds (in the sense of~\cite{rainer}) form a basis for the topology
on $M$.  Thus, we may choose regular diamonds $O_{3},O_{4}$ such that
$\overline{O_{3}}\subseteq O_{1}$ and $\overline{O_{4}}\subseteq O_{2}$.  The
nonfiniteness of the local algebras $\alg{R}_{\pi}(O_{3}),\alg{R}_{\pi}(O_{4})$ is established
in~\cite[Thm.~3.6.g]{rainer}, and the split property for these algebras is established
in~\cite[Thm.~3.6.d]{rainer}.  Since the split property entails the Schlieder property, it follows
from Proposition~1 that the set of vectors inducing Bell correlated states for
$\alg{R}_{\pi}(O_{3}),\alg{R}_{\pi}(O_{4})$ [and thereby Bell correlated for
$\alg{R}_{\pi}(O_{1}),\alg{R}_{\pi}(O_{2})$] is dense in $\hil{S}_{\pi}$.  \end{proof}

There are many physically interesting states, such as the Minkowski vacuum itself, about which
Propositions~3 and~4 are silent.  However, Reeh-Schlieder type theorems entail that many of
these physically interesting states are induced by vectors which are cyclic for local algebras, and
thus it follows from Proposition~2 that these states are nonseparable across any spacelike
separated pair of local algebras.  In particular, although there is an upper bound on the Bell
correlation of the Minkowski vacuum (in models with a mass gap) that decreases exponentially
with spacelike separation~\cite[Prop.~3.2]{invariants}, the vacuum state remains nonseparable (in
our sense) at all distances.  On the other hand, since nonseparability is only a \emph{necessary}
condition for Bell correlation, none of our results decide the question of whether the vacuum
state always retains \emph{some} Bell correlation across arbitrary spacelike separated regions.     

\begin{center} {\bf Acknowledgments} \end{center} 
We are extremely grateful to David Malament (UC Irvine) for encouraging our work on this
project.  We wish also to thank the anonymous referee for helping us to understand the scope of
our results (which led to significant improvements in section IV), and Rainer Verch
(G{\"o}ttingen) for providing helpful correspondence.

\end{document}